\documentclass[10pt,aps,floats,prd,
twocolumn,showpacs]{revtex4}
\usepackage{graphicx}
\usepackage{epsfig}

\newcommand{\beq}{\begin{equation}}
\newcommand{\eeq}{\end{equation}}
\newcommand{\be}{\begin{eqnarray}}
\newcommand{\ee}{\end{eqnarray}}

\def\+{\dagger}

\begin{document}
\title{Width of the 511~KeV Line from the Bulge of the Galaxy }
\author{  Ariel  Zhitnitsky}

\affiliation{ Department of Physics and Astronomy, University of British 
Columbia, Vancouver, BC, V6T 1Z1, CANADA}
%\date{Nov 15.2006}

\begin{abstract}
In this paper  I  present the detail estimations  
for the width
 of the  511~  keV  line produced by   a mechanism when
 dark matter is represented by macroscopically large dense
 nuggets. I argue that the width of 511 keV emission  in this mechanism   is very narrow ( in a few keV range)
  in agreement with all observations.
The dominant mechanism of the annihilation  in this case is the 
positronium  formation $ e^+e^-\rightarrow  ~ ^1S_0 \rightarrow 2\gamma$ rather than  a direct 
 $e^+e^-\rightarrow 2\gamma$ annihilation.  
I also discuss some generic features of the $\gamma$ rays spectrum (in few MeV range) resulting from
this mechanism. 
\end{abstract}
\pacs{98.80.Cq, 95.30.Cq, 95.35.+d, 12.38.-t}
\maketitle

%{\em Introduction}.
\section{Introduction}
Recent observations of the galactic centre have presented a number of puzzles 
for our current understanding of galactic structure and astrophysical 
processes. In particular a series of independent observations 
have detected an excess flux of  photons across a broad 
range of energies. Specifically, 
 SPI/INTEGRAL observations of the galactic centre have 
detected an excess of 511 keV gamma rays resulting 
from low momentum electron-positron annihilations. The observed intensity 
  is a mystery.    After accounting for known positron
sources, only a small fraction of the emission may be
explained~\cite{Knodlseder:2003sv,Jean:2003ci,Boehm:2003bt,
Beacom:2005qv,Zhang:2006fr,Yuksel:2006fj}.  
   Motivated by this observation, it has been suggested  recently \cite{Oaknin:2004mn} 
   that the observed flux can be   explained by the  idea 
that dark matter (DM)  particles are
strongly interacting composite macroscopically large objects which made of
 well known light quarks or even antiquarks \cite{qcdball}\cite{Oaknin:2003uv}, similar 
to the Witten's strangelets \cite{Witten:1984rs}.  

The width of 511 keV line has been measured  by  SPI/INTEGRAL on the level of few keV~\cite{Knodlseder:2003sv,Jean:2003ci},
but has not been calculated in the original paper \cite{Oaknin:2004mn}.
The goal of the present work is to fill this gap.   More precise, 
    in the present  paper I  will estimate
    the  probability for the positronium formation and  
     argue that the positronium formation (rather than direct
      annihilation) plays a 
      dominant role in $e^+e^-$ annihilation when an electron from visible matter hits
      the antimatter nugget.  In this case the estimated width of 511 keV line  
      is determined by the velocity distribution of the positroniums which move with typical velocities $v\sim \alpha$, see section III.
  Consequently, this motion  determines  the  width of $511 $ KeV line to be $\Gamma\sim m_e\alpha \sim $ few keV 
      in agreement with measurements. The direct annihilation $e^+e^-\rightarrow 2\gamma$ which lead to the continuum spectrum 
      with typical   photons in MeV range 
      is a sub leading
      process as we argue below. This   direct annihilation $e^+e^-\rightarrow 2\gamma$ might be interesting on its own
      as it may explain a well known mystery on
     the excess of gamma-ray photons detected by COMPTEL in $\sim 1-20$ MeV  energy range.
However, the corresponding analysis is the  
      subject of a different
      paper \cite {Lawson:2007kp} and shall not be discussed here. 

 \section{ Compact Composite Objects (CCO) }

Unlike conventional dark matter candidates, dark matter/antimatter nuggets
are strongly interacting,  macroscopically large objects\cite{Zhitnitsky:2006vt}.  Such a ``counterintuitive" 
proposal does  not contradict any of the many known observational constraints on
dark matter or antimatter in our universe due to three main reasons: 
  1) the nuggets carry a huge (anti)baryon charge $|B| \approx
10^{20}$ -- $10^{33}$, so they have a macroscopic size and  a tiny number density.  2) They
have   nuclear densities in the bulk, so their interaction cross-section per unit mass is small
$\sigma/M \approx 10^{-13}$ -- $10^{-9}$~cm$^2$/g. This small factor effectively
replaces a  condition on weakness of  interaction of  conventional dark matter candidates such as 
WIMPs.
3) They have a large binding energy (gap
$\Delta \approx 100$~MeV) such that baryons in the nuggets are not
available to participate in big bang nucleosynthesis (BBN) at $T
\approx 1$~MeV\@.  Therefore,   CCOs  do not contribute
to  $\Omega_B$, but rather, they   do   contribute to the 
``non-baryonic" cold dark matter $\Omega_{DM}$ of the universe.
 On large scales, the CCOs are sufficiently dilute
that they behave as standard collisionless cold dark matter.
As we mentioned above,
 CCOs can be made from matter as well as from antimatter.  
 Precisely these nuggets   made of antimatter represent an  unlimited  source of positrons which  
 can annihilate with visible electrons and produce observed photons.

 For our purposes we adopt  a simple model for a
  Compact Composite Object when all 
  quarks form one of the  color super conducting (CS) phases
  with densities few times the nuclear density \cite{cs_r}, while  the electrons   in CCOs
 can be treated as noninteracting fermi gas with density $n_e\simeq \frac{(\mu^2-m_e^2)^{3/2}}{3\pi^2}$,
  with $\mu$ being the electron chemical potential. A numerical estimation of 
$\mu$  strongly depends on the specific details  of CS phase under consideration,
 and varies  from few MeV to tens (or even hindred) MeV,~\cite{cs_r}- \cite{olinto}.  
It is also assumed that the nuggets have very thin electro- sphere with a
``transition region"  of a  microscopical scale separating  the bulk of the dense matter (with large $\mu$) from vacuum (with $\mu=0$).
There existence of this ``transition region"  is a very generic feature of the system and is the direct consequence of the Maxwell's equations and 
  the chemical equilibrium requirement \cite{olinto}.

 Our goal here is to argue that the  photon spectrum  resulting from 
CCO- based mechanism  of $e^+e^-$ annihilation has  the following main features:
            The dominant fraction of incoming electrons   will form positroniums. 
As is known, once the  positroniums are formed,
    one quarter of them
        (in $^1S_0$ state) will 
     eventually  decay   to two   511~ keV photons, while three quarters
     of them (in $^3S_1$ states)   will produce a continuum with the typical energies in 
     100~  keV range.  A small fraction of electrons will experience  the direct annihilation  $e^+e^-\rightarrow 2\gamma$.  
     The typical photons produced in direct annihilation will have  energies of order $\mu\sim $ few MeV. 
     Such photons from the direct annihilation must always accompany the 511 keV line as they 
   produced by the same mechanism within our framework.  We shall not discuss 
    the spectrum and intensity of the $\sim 1-20~$ MeV  gamma-rays in the present paper referring to  \cite {Lawson:2007kp}.
    However, we would like to remark here that the excess of photons measured by COMPTEL precisely in this band, $\sim 1-20~$ MeV\cite{Strong}
    can be naturally explained by this mechanism if one assumes that the fraction of incoming electrons 
   (which avoid  the positronium formation and   can reach the nugget's surface with large $\mu$ ) is on the level of $\sim 10 \%$
   while  the dominant fraction of incoming electrons   $\sim 90 \%$ will form positroniums \cite {Lawson:2007kp}.

     Important remark here is that $\mu$ is always in MeV region, much larger than the typical atomic energy scale which is in  eV range. In this case the results which follow
are not very 
sensitive to the specific properties of CS phase in the bulk. Therefore, our simplified treatment of the leptons
 as noninteracting fermi gas  
is sufficiently good  approximation for this problem: any changes (due to the interactions in the bulk of nuggets)
   are happening  at the $\mu \sim $ MeV scale.
These changes  do not affect physics on eV  scale which is the subject of this paper.
 
        \section{Positronium Formation }
         We now consider the probability for the positronium formation when 
             electrons   hit the CCO made of antimatter.
  What is the fate of these non relativistic   electrons? We shall argue below that the most
  likely outcome of these events is a formation of 
  the  bound states (positroniums with arbitrary quantum numbers $|n,l,m\rangle$)
    which eventually decay to  two photons  with $\hbar\omega \simeq $511 keV  and with width
    $\Gamma\sim m_e\alpha \sim$ few keV 
    or three photons with well known continuum spectrum $0< \hbar\omega < m_e$.    
     
     Indeed, consider a system  of an incoming electron   and a positron  from a nugget  with momenta $\vec{q_1}$ and $\vec{q_2}$ correspondingly.
     Assuming that both particles are non relativistic
     %\footnote{Electron's velocity $v_e\sim 10^{-3}c$
     %is always assumed to be small, while the positron's velocity has to be small for the probability
     %of the positronium formation to be order of one, see below.},
     we can calculate the probability of positronium formation  with   quantum numbers $|n,l,m\rangle$
     by expanding the original wave functions (plane waves with momenta $\vec{q_1}$ and $\vec{q_2}$)
     in terms of the new basis of   positronium's bound states (plus continuum), 
     \beq
     \label{psi}
     \Psi_{q_1, q_2} (r_1, r_2)=e^{iQR}\sum_{nlm}c_{nlm}(q)\psi_{nlm}(r) +cont. ,
     \eeq
     where $r\equiv (r_1-r_2), ~q\equiv 1/2(q_1-q_2)$ correspond to 
      the relative coordinate and momenta,
     while $R\equiv 1/2(r_1+r_2), ~Q\equiv (q_1+q_2)$  describe the   
     center of mass of the $e^+e^-$ system.
     By definition, $|c_{nlm}(q)|^2$ gives the  probability to find $e^+e^-$ system in the positronium 
     state with quantum numbers $|n,l,m\rangle$ if initial $e^+e^-$ states  had momenta $\vec{q_1}$ and $\vec{q_2}$  with proper normalization. In particular, for the ground state,
     \beq 
     \label{c_0}
  |c_{100}(q)|^2\sim  |\int e^{-r/a}e^{i\vec{q}\vec{r}}d^3r|^2\sim (a^2q^2+1)^{-4},  
   \eeq
where  $a\equiv 1/(m\alpha)\simeq 10^{-8}cm$ is the Bohr radius. 

Few remarks now are in order:\\
a) The probability for the positronium formation is large when  $q$ is sufficiently small, $q \sim 1/a\sim m\alpha$. 
This justifies our treatment of positrons 
 as nonrelativistic particles. In different words, a non relativistic incoming electron will pick up a positron from Fermi gas with a small (rather than   large) momenta $q\sim m\alpha$ to form a positronium.
 Probability of formation of the positroniums with large $q\gg a^{-1}$ is exceedingly small.
 \\
b) The expression for the probability  of the positronium formation 
 does not contain a
small factor $\alpha^2$ which is inherent feature of the direct annihilation process, see below eq.(\ref{sigma1});\\
c) Once positroniums are formed, they will eventually decay much later (within or outside CCO)
to two/ three photons  producing the low energy spectrum discussed above: 511 keV line + well known continuum spectrum $0< \hbar\omega < m_e$; \\
d) One may wonder why a small coupling constant $\alpha^2$ does not enter the expression
for the process which eventually leads to the photon's emission.
The answer  of course is related to the resonance nature of the phenomena.
Similar situation occurs, e.g. in charge exchange processes such as capture of an electron 
 from a hydrogen atom by a slow moving proton;\\
 e) The fact that the positronium formation plays a crucial role  in the theory of positron annihilation in solids, has been known for 50 years\cite{old}, see also recent review on the subject\cite{new}.
  Positronium formation always takes place whenever  it is energetically allowed and velocities are small
 ( when the so called ``Ore gap" is not destroyed by a complex condensed matter system);\\
 f) The magnitude of width of 511 KeV line in our framework 
 is determined by the velocity distribution of the positroniums. 
 Indeed, the positroniums in our framework are formed not at rest,
 but instead they carry a   nonzero momenta $Q\equiv (q_1+q_2)$ as eq.(\ref{psi}) suggests.
As was  argued above, parametrically 
 $Q\sim a^{-1}\sim m_e\alpha$. It implies that once the  positroniums are formed, they do 
 move with typical  velocities $v\sim Q/m_e\sim \alpha $. Consequently, this motion  leads to the  width of $511 $ KeV line to be $\Gamma\sim m_e\alpha \sim $ few keV due to the Doppler  effect;\\
 g) The probability for the positronium formation  
 is order of one for  small $q$ as follows from eqs.(\ref{psi},\ref{c_0}). However, these equations
 do  not say what is the time scale   saturating  this large probability.
    
 Therefore, 
 the crucial question is:   what is the time scale $\tau_{Ps}$ for the positronium formation in our specific circumstances?
 If this time scale is sufficiently short, then  an incident  electron has a great chance to form 
 the positronium (which eventually leads to 511 keV line) before it reaches the nugget's surface where the typical positron energies are large $\sim\mu$. If, on other hand,
 this time scale is very large, then  an  incident  electron
 very likely will reach the surface of the nugget and will experience the direct $e^+e^-\rightarrow 2\gamma$ annihilation
 with emission of $\sim $ MeV photons.
% If this time scale is much longer than  the time scale $ \tau_{e^+e^-}$ calculated above
% (\ref{tau1}),  the bound states would not have time to   form, and positronium formation
% would not take place. In opposite case, $ \tau_{e^+e^-} \gg \tau_{Ps}$ 
% the electrons will have a plenty of time 
 %to form positroniums before the direct annihilation (\ref{tau1}) happens.
  
  The cross section for the resonance positronium formation in atomic units is order of one.
         In conventional units it is $\sigma (e^+e^-\rightarrow Ps)\sim a^2$\footnote{If relative velocity $v_e$ of colliding $e^+e^-$ particles is much smaller than 
         atomic velocity $\sim \alpha$ there will be an additional enhancement factor $\alpha/v_e$
         in the cross section due to the long distance Coulomb interactions\cite{Landau}.}.  In order to estimate 
          $\tau_{Ps}^{-1}$ we have to multiply $ \sigma (e^+e^-\rightarrow Ps)$ by
          density of positrons which  effectively participate  in the positronium formation 
          and atomic velocity
          which is order of $v\sim \alpha$. 
          
          The density of positrons surrounding the anti matter nugget can be easily estimated 
          in the transition region by using the Thomas- Fermi (mean field) -like approximation~\cite{olinto}. In the relativistic 
          regime the density behaves like $n(z)\sim 1/z^3$ where $z$ is the distance from the 
          nugget's surface~\cite{olinto}. One can show that this behavior slowly changes to $n(z)\sim a^3/z^6$
          in non relativistic regime where $a\sim (m\alpha)^{-1}$ is  the Bohr radius. 
          We do not  need to know an exact numerical coefficients in this formula. The  important thing
         for our discussions in what follows  is there existence of a  transition region (``electro -sphere"~\cite{olinto}) where chemical potential   $\mu$   interpolates
       between a large value on the surface of the nugget and zero value far away from the nugget. This interpolation always
       includes a region with a  typical atomic density
       $n\sim a^{-3}$ at distance $z\sim a\sim 10^{-8}$cm  from the nugget's surface. 
          
           Collecting all these factors together we arrive to the following estimation
           for the probability $P$ that incident electron entering the ``electro -sphere" of the nugget will form positronium 
         \be
 \label{Ps1}
  \tau_{Ps}^{-1} \equiv \frac{dP}{dt}  \sim v \cdot \sigma (e^+e^-\rightarrow Ps) \cdot n(z\sim a)\sim 
  v/a ,
  \ee
  where we use $n\sim a^{-3}$ for $z\sim a$. This expression clearly shows that
  the total probability  for the positronium formation (which  consequently decay producing 511 keV line)
  becomes of order of one  at atomic  distances $z\sim a$ from the nugget's surface, i.e. long before the incident electron reaches   
 the region of large  positron densities close to the nugget's surface.

 This result is  in a clear contrast with estimations presented in\cite{CSS}
  where a MeV -broad  spectrum is predicted resulting from the same, CCO-based mechanism.
  The crucial ingredient in our estimates  is of course, the resonance behavior  for 
  the cross section $\sigma (e^+e^-\rightarrow Ps)\sim a^2\sim (m\alpha)^{-2}$  
  in contrast with non resonance formula   for the direct $e^+e^-\rightarrow 2\gamma$ annihilation when small parameter $\alpha^2$ enters
  the numerator in the corresponding formula, see e.g. \cite{Peskin},
    \beq
   \label{sigma1}
   \sigma(e^+e^-\rightarrow 2\gamma )\simeq \frac{2\pi\alpha^2}{s}\ln\left(\frac{s}{4m_e^2}\right), ~~ s \gg 4m_e^2.
      \eeq

  To conclude this section: the dominant portion of all electrons 
    falling into CCO (made of antimatter) will form the positroniums which eventually decay with low energy spectrum described above.
      The  typical width
  of  outgoing flux of $511 KeV$ photons is of order $\Gamma\sim \alpha m\sim $ few KeV.
  These features are very universal and  do not depend on  specific details  of the  nugget's  internal structure (such as a large variation of possible CS phases
  in the bulk). 
    Some incident electrons entering the nugget's surface will experience the direct annihilation with emission of     gamma-rays in MeV band.
    These photons, may even have been already observed\cite {Lawson:2007kp}.     However, the direct annihilation
      plays a sub leading  role as argued above.

  \section{ Conclusion.}
 We   present a generic picture 
  of the $\gamma$ spectrum  
   which results from the CCO -based mechanism. 
As we argued above,  the vast majority of $e^+e^-$ annihilations go through the
 positronium formation with the width of $511 $KeV line  to be $\Gamma\sim m_e\alpha
 \sim$ few KeV. This is precisely what has been observed {\cite{Knodlseder:2003sv},\cite{spectrum1},\cite{spectrum2}.
 Also: this line is always accompanied by  
 the well-known
           continuum with energies $\hbar \omega < 511~ $KeV from  the  $^3S_1$      positronium decays
           (with the ratio $3$ to $1$).
           Amazingly, this is precisely the spectrum 
           obtained in the recent analysis   with fraction of the observed 
           positroniums  estimated to be  $(96.7\pm 2.2)\%$\cite{spectrum1},
           $(92\pm 9)\%$ \cite{spectrum2}. Undoubtedly, these observations are
             consistent with  almost $\simeq 100\%$  
           positronium fraction  predicted by CCO based mechanism due to the 
           strong suppression of the direct annihilation in the region $\hbar \omega \leq  511~ $KeV. 
           
           Our mechanism also suggests that the 511 KeV line must be accompanied by very broad
           (1 -20 ~MeV) spectrum with the spectral density $\frac{d\Phi}{d\omega} $ at 
        $\hbar \omega \simeq 511~ $KeV  few orders
        of magnitude smaller 
        than from the positronium decays. However, the
         total integrated  flux over the large region 
       $\int ^{\mu}_0\frac{d\Phi}{d\omega}d\omega$ could be sufficiently large. 
         Amazingly, there is indeed an  observational 
       evidence for an
       excess of photons in (1 -20) ~MeV  region, see 
        \cite{Strong}, \cite{MeV}
       and references on the original works therein.
       It has been   argued that  the soft gamma-ray spectrum in (1 -20) ~MeV  region cannot fully be attributed to either Active Galactic Nuclei or Type Ia supernovae or a combination of the two \cite{MeV}.
  Therefore,   the (1 -20) ~MeV  observed excess  may find its natural explanation 
  as a result of the direct annihilation of visible electrons with CCO's positrons. 
   Such an explanation can be confirmed (or ruled out)    
  if the correlation between        (1 -20) ~MeV   photons and   511 KeV line is established.
   We shall not discuss 
   this problem  in the present paper by referring to  \cite {Lawson:2007kp}.
    However, we would like to remark here that the excess of photons measured by COMPTEL  
        can be naturally explained by this mechanism \cite {Lawson:2007kp} if one assumes that a small fraction ( on the level of $\sim 10 \%$) of incoming electrons 
  can avoid  the positronium formation and   can reach the nugget's surface where $\mu$ is large. 
       
  Similar correlation should also exist between  511 KeV line and diffuse X ray emission
  as discussed in details in \cite{Forbes:2006ba}.

 % \section*{Acknowledgements}

 I am thankful to I. Khriplovich for the discussions during his visit to Vancouver.
 This work was supported in part by the National Science and Engineering
Research Council of Canada.


\section*{References}
  \begin{references}
  
   
\bibitem{Knodlseder:2003sv}
J.~{Kn\"odlseder} {\em et~al.},
\newblock Astron. Astrophys. {\bf 411}, L457 (2003), arXiv:astro-ph/0309442.
%%CITATION = ASTRO-PH 0309442;%%
%\cite{Jean:2003ci}
\bibitem{Jean:2003ci}
P.~Jean {\it et al.},
%``Early SPI/INTEGRAL measurements of galactic 511 keV line emission from
%positron annihilation,''
Astron.\ Astrophys.\  {\bf 407}, L55 (2003).
[arXiv:astro-ph/0309484].
%%CITATION = ASTRO-PH 0309484;%%

%\cite{Boehm:2003bt}
\bibitem{Boehm:2003bt}
C.~Boehm, D.~Hooper, J.~Silk and M.~Casse,
%``MeV dark matter: Has it been detected?,''
Phys.\ Rev.\ Lett.\  {\bf 92}, 101301 (2004).
%[arXiv:astro-ph/0309686].

\bibitem{Beacom:2005qv}
J.~F. Beacom and H.~Yuksel,
\newblock Phys. Rev. Lett. {\bf 97}, 071102 (2006), arXiv:astro-ph/0512411.
%%CITATION = ASTRO-PH 0512411;%%

\bibitem{Zhang:2006fr}
  L.~Zhang, X.~L.~Chen, Y.~A.~Lei and Z.~G.~Si,
  %``The impacts of dark matter particle annihilation on recombination and  the
  %anisotropies of the cosmic microwave background,''
  Phys.\ Rev.\  D {\bf 74}, 103519 (2006)
  [arXiv:astro-ph/0603425].
  
\bibitem{Yuksel:2006fj}
H.~Yuksel,
\newblock (2006), arXiv:astro-ph/0609139.
%%CITATION = ASTRO-PH 0609139;%%

 \bibitem{Oaknin:2004mn}
  D.~H.~Oaknin and A.~R.~Zhitnitsky,
  %``511-keV photons from color superconducting dark matter,''
  Phys.\ Rev.\ Lett.\  {\bf 94}, 101301 (2005)
 % [arXiv:hep-ph/0406146].

  \bibitem{qcdball}
 %\bibitem{Zhitnitsky:2002qa}
  A.~R.~Zhitnitsky,
  %``'Nonbaryonic' dark matter as baryonic color superdonductor,''
  JCAP {\bf 0310}, 010 (2003)
 % [arXiv:hep-ph/0202161].%\cite{Zhitnitsky:2002nr}
%\item%{Zhitnitsky:2002nr}
%{\it ``Dark matter as dense color superconductor''}
%\\{}A.~R.~Zhitnitsky,
%~arXiv:astro-ph/0204218,~
%\href{http://www.slac.stanford.edu/spires/find/hep/www?eprint=astro-ph\%2F0204$  
%Nucl.\ Phys.\ Proc.\ Suppl.\  \ B{ \bf 124}, 99 (2003); J.\ Phys.\ G {\bf 30}, S513 (2004).
\bibitem{Oaknin:2003uv}
 D.~H.~Oaknin and A.~Zhitnitsky,
  %``Baryon asymmetry, dark matter and quantum chromodynamics,''
  Phys.\ Rev.\ D {\bf 71}, 023519 (2005).
 % [arXiv:hep-ph/0309086].

\bibitem{Witten:1984rs}
E.~Witten,
%``Cosmic Separation Of Phases,''
Phys.\ Rev.\ D {\bf 30}, 272 (1984).
 
 
\bibitem{Lawson:2007kp}
  K.~Lawson and A.~R.~Zhitnitsky,
  %``Diffuse cosmic gamma-rays at 1-20 MeV: A trace of the dark matter?,''
  arXiv:0704.3064 [astro-ph].
  %%CITATION = ARXIV:0704.3064;%%
  \bibitem{Strong}
  A.~W.~Strong, I.~V.~Moskalenko and O.~Reimer,
    %``Diffuse continuum gamma rays from the Galaxy,''
      Astrophys.\ J.\  {\bf 537}, 763 (2000)
      [Erratum-ibid.\  {\bf 541}, 1109 (2000)]
      [arXiv:astro-ph/9811296],

\bibitem{Zhitnitsky:2006vt}
A.~Zhitnitsky,
\newblock Phys. Rev. {\bf D74}, 043515 (2006), arXiv:astro-ph/0603064.
%%CITATION = ASTRO-PH 0603064;%%


\bibitem{cs_r}
K.~Rajagopal and F.~Wilczek, 
arXiv:hep-ph/0011333;
M.~G.~Alford, Ann.\ Rev.\ Nucl.\ Part.\ Sci.\  {\bf 51}, 131 (2001);
T.~Schafer and E.~V.~Shuryak, Lect.\ Notes Phys.\  {\bf 578}, 203 (2001)
 \bibitem{Steiner:2002gx}
A.~W.~Steiner, S.~Reddy and M.~Prakash,
%``Color-neutral superconducting quark matter,''
Phys.\ Rev.\ D {\bf 66}, 094007 (2002).

\bibitem{Madsen:2001fu}J.~Madsen,
%``Color-flavor locked strangelets,''
Phys.\ Rev.\ Lett.\  {\bf 87}, 172003 (2001)
%[arXiv:hep-ph/0108036].
%%CITATION = HEP-PH 0108036;%%

 \bibitem{olinto} C.~Alcock, E.~Farhi and A.~Olinto, 
Astrophys.\ J.\  {\bf 310}, 261 (1986).

 
\bibitem{old} R. Ferrell, Review of Modern Physics, vol {\bf 28}, 308 (1956).
\bibitem{new} M. Puska and R. Nieminen, Review of Modern Physics, vol {\bf 66}, 841 (1994).


\bibitem{CSS}
%\bibitem{Cumberbatch:2006bj}
  D.~T.~Cumberbatch, J.~Silk and G.~D.~Starkman,
  %``Difficulties for Compact Composite Object Dark Matter,''
  arXiv:astro-ph/0606429.

\bibitem{Peskin} M. Peskin and D. Schroeder, " An Introduction to QFT", Addison-Wesley Publishing Company, 1996.

  \bibitem{Landau} L.D.Landau and E.M.Lifshitz, Quantum Mechanics, Volume III, Elsevier Butterworth-Heinemann, 2005. 
 \bibitem{spectrum1}
P.~Jean {\it et al.}, astro-ph/0509298
\bibitem{spectrum2}
G. Weidenspointner  {\it et al.}, astro-ph/0601673
  \bibitem{MeV} K. Ahn, E. Komatsu and P. Hoflich, Phys. Rev. D {\bf 71}, 121301(R) (2005);
  K. Ahn, E. Komatsu, Phys. Rev. D {\bf 72}, 061301(R) (2005);
 L. E. Strigari, J. F. Beacom, T. P. Walker and P. Zhang, J. Cosmol. Astropart. Phys. {\bf 04}  (2005) 017. 

\bibitem{Forbes:2006ba}
  M.~M.~Forbes and A.~R.~Zhitnitsky,
  %``Dark antimatter as a galactic heater: X-rays from the core of our galaxy,''
  arXiv:astro-ph/0611506.
  %%CITATION = ASTRO-PH/0611506;%%

 \end{references}
\end{document}